\documentclass[right=1.2in]{article}

\usepackage{mystyle}

\usepackage[utf8]{inputenc} % allow utf-8 input
\usepackage[T1]{fontenc}    % use 8-bit T1 fonts
\usepackage{hyperref}       % hyperlinks
\usepackage{url}       % simple URL typesetting
\usepackage{xcolor}
\usepackage{booktabs}       % professional-quality tables
\usepackage{amsfonts}       % blackboard math symbols
\usepackage{nicefrac}       % compact symbols for 1/2, etc.
\usepackage{microtype}      % microtypography
\usepackage{cleveref}       % smart cross-referencing
\usepackage{lipsum}         % Can be removed after putting your text content
\usepackage{graphicx}
\usepackage{natbib}
\usepackage{subcaption}
\usepackage{doi}
\usepackage{multirow}
\usepackage{array}
\usepackage{geometry}
\usepackage{rotating}
\usepackage{array}

\title{The relationship between episcopal genealogy and ideology in the Roman Catholic Church}

\usepackage{authblk}

\newbox{\orcid}\sbox{\orcid}{\includegraphics[scale=0.06]{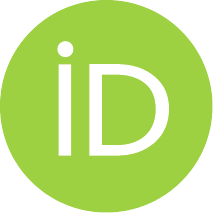}}

\author[1]{%
	\href{https://orcid.org/0000-0000-0000-0000}{\usebox{\orcid}\hspace{1mm}
		Marta Baratto%\thanks{\texttt{marta.baratto@unito.it}}
	}%
}
\author[2,3]{%
	\href{https://orcid.org/0000-0000-0000-0000}{\usebox{\orcid}\hspace{1mm}Ivan Casanovas%\thanks{\texttt{ivan.casanovas@ub.edu}}
	}%
}
\author[4]{%
	\href{https://orcid.org/0000-0000-0000-0000}{\usebox{\orcid}\hspace{1mm}
		Ivan Decostanzi%\thanks{\texttt{ivan.deconstanzi@isi.it}}
	}%
}
\author[5]{%
	\href{https://orcid.org/0000-0000-0000-0000}{\usebox{\orcid}\hspace{1mm}Henrique Machado Borges%~\thanks{\texttt{h.machadoborges@northeastern.edu}}
	}%
}
\author[6]{%
	\href{https://orcid.org/0009-0000-2024-2427}{\usebox{\orcid}\hspace{1mm}
		Samuel Martínez Alcalá%\thanks{\texttt{samuel.martinez@mncn.csic.es}}
	}%
}
\author[7,8]{%
	\href{https://orcid.org/0000-0000-0000-0000}{\usebox{\orcid}\hspace{1mm}Ilaria Stanzani%\thanks{\texttt{ilaria.stanzani@edu.unige.it}}
	}%
}
\author[9]{%
	\href{https://orcid.org/0000-0000-0000-0000}{\usebox{\orcid}\hspace{1mm}
		Alberto Antonioni%\thanks{\texttt{alberto.antonioni@uc3m.es}}
	}%
}
\author[10]{%
	\href{https://orcid.org/0000-0001-8794-6410}{\usebox{\orcid}\hspace{1mm}
		Iacopo Iacopini%\thanks{\texttt{i.iacopini@northeastern.edu}}
	}%
}
\author[11]{%
	\href{https://orcid.org/0000-0002-1074-0411}{\usebox{\orcid}\hspace{1mm}
		Michele Re Fiorentin\thanks{\texttt{michele.refiorentin@polito.it}}}%
}
\author[12]{%
	\href{https://orcid.org/0000-0002-9246-6195}{\usebox{\orcid}\hspace{1mm}
		Eugenio Valdano%\thanks{\texttt{eugenio.valdano@inserm.fr}}
	}%
}

\affil[1]{Physics Department, Università degli Studi di Torino, Torino, Italy}
\affil[2]{Departament de Física de la Matèria Condensada, Universitat de Barcelona, Martí i Franquès, 1, Barcelona, Spain}
\affil[3]{Universitat de Barcelona Institute of Complex Systems, Martí i Franquès, 1, Barcelona, Spain}
\affil[4]{ISI Foundation, Turin 10126, Italy}
\affil[5]{NP Lab, Network Science Institute, Northeastern University, London, E1W 1LP, UK}
\affil[6]{Departamento de Biodiversidad y Biología Evolutiva, Museo Nacional de Ciencias Naturales (CSIC), C/ José Gutiérrez Abascal, 2. 28006. Madrid, Spain}
\affil[7]{Dipartimento di Informatica, Bioingegneria, Robotica e Ingegneria dei Sistemi, Università di Genova, Genova, Italy}
\affil[8]{Machine Learning Genoa Center, Università di Genova, Genova, Italy}
\affil[9]{GISC, Department of Mathematics, Carlos III University of Madrid, 28911 Leganés, Spain}
\affil[10]{Network Science Institute, Northeastern University London, London, E1W 1LP, UK}
\affil[11]{Department of Applied Science and Technology, Politecnico di Torino, c. Duca degli Abruzzi 24, 10129 Torino, Italy}
\affil[12]{Sorbonne Université, INSERM, Institut Pierre Louis d’Epidémiologie et de Santé Publique, F75012, Paris, France}

\hypersetup{
	pdftitle={The association between genealogical and ideological proximity in the Roman Catholic Church},
	pdfsubject={},
	pdfauthor={Baratto et al.},
	pdfkeywords={Topic modeling,  genealogy networks, Roman Catholic Church, opinion polarization, cardinal, bishop, clergy, consecration},
}

\begin{document}
	\maketitle
	
	\begin{abstract}

		In this study we investigate how hierarchical structures within the Roman Catholic Church shape the ideological orientation of its leadership. The full episcopal genealogy dataset comprises over 35,000 bishops, each typically consecrated by one principal consecrator and two co-consecrators, forming a dense and historically continuous directed network of episcopal lineage. Within this broader structure, we focus on a dataset of 245 living cardinals to examine whether genealogical proximity correlates with doctrinal alignment on a broad set of theological and sociopolitical issues. We identify motifs that capture recurring patterns of lineage, such as shared consecrators or co-consecrators. In parallel, we apply natural language processing techniques to extract each cardinal’s publicly stated positions on ten salient topics, including LGBTQIA+ rights, women’s roles in the Church, liturgy, bioethics, priestly celibacy, and migration. Our results show that cardinals linked by specific genealogical motifs, particularly those who share the same principal consecrator, are significantly more likely to exhibit ideological similarity. We find that the influence of pope John Paul II persists through the bishops he consecrated, who demonstrate systematically more conservative views than their peers. These findings underscore the role of hierarchical mentorship in shaping ideological coherence within large-scale religious institutions. Our contribution offers quantitative evidence that institutional lineages, beyond individual background factors, may have an impact on the transmission and consolidation of doctrinal positions over time.

		%In this study we investigate how bishop consecrations shape ideological alignment within the Roman Catholic Church through a unique analysis of 245 living Catholic cardinals, examining whether shared mentorship lineages influence doctrinal positions on ten traditional and contemporary issues. Using network science and natural language processing, we map the ecclesiastical genealogy connecting cardinals through episcopal consecrations and analyzed their stated positions on topics such as LGBTQIA+ rights, women's ordination, celibacy, and bioethics.
		
		%Cardinals who share mentorship lineages—particularly those consecrated by the same principal consecrating bishop—exhibit significantly more similar ideological positions than expected by chance. 
		%Network analysis reveals cardinals are more interconnected than random chance would predict, forming a tight-knit hierarchical structure. 
		%Despite this connectivity, clear ideological polarization emerges on most issues.
		
		%These findings demonstrate that institutional mentorship networks systematically influence belief alignment beyond demographic factors, providing quantitative evidence for how social transmission shapes ideological coherence in large-scale organizations.

	\end{abstract}

	% keywords can be removed
	
	%\section{Authors Order}
	%Participants in Alphabetical Order + Tutors at the end
	
	\section{Introduction}
	%Why we want to do this? 
	%what are we looking for? 
	%why is it interesting? 
	%Who has done something similar? 
	
	% IDEA for the structure of the introduction
	
	% Introduction/Background to our problem
	% Our problem/data: network and textual data
	% Already existing techniques:
	% Topic modelling
	% Network 
	% Our proposal
	
\begin{figure}[b]
		\centering
		\includegraphics[width=0.8
		\textwidth]{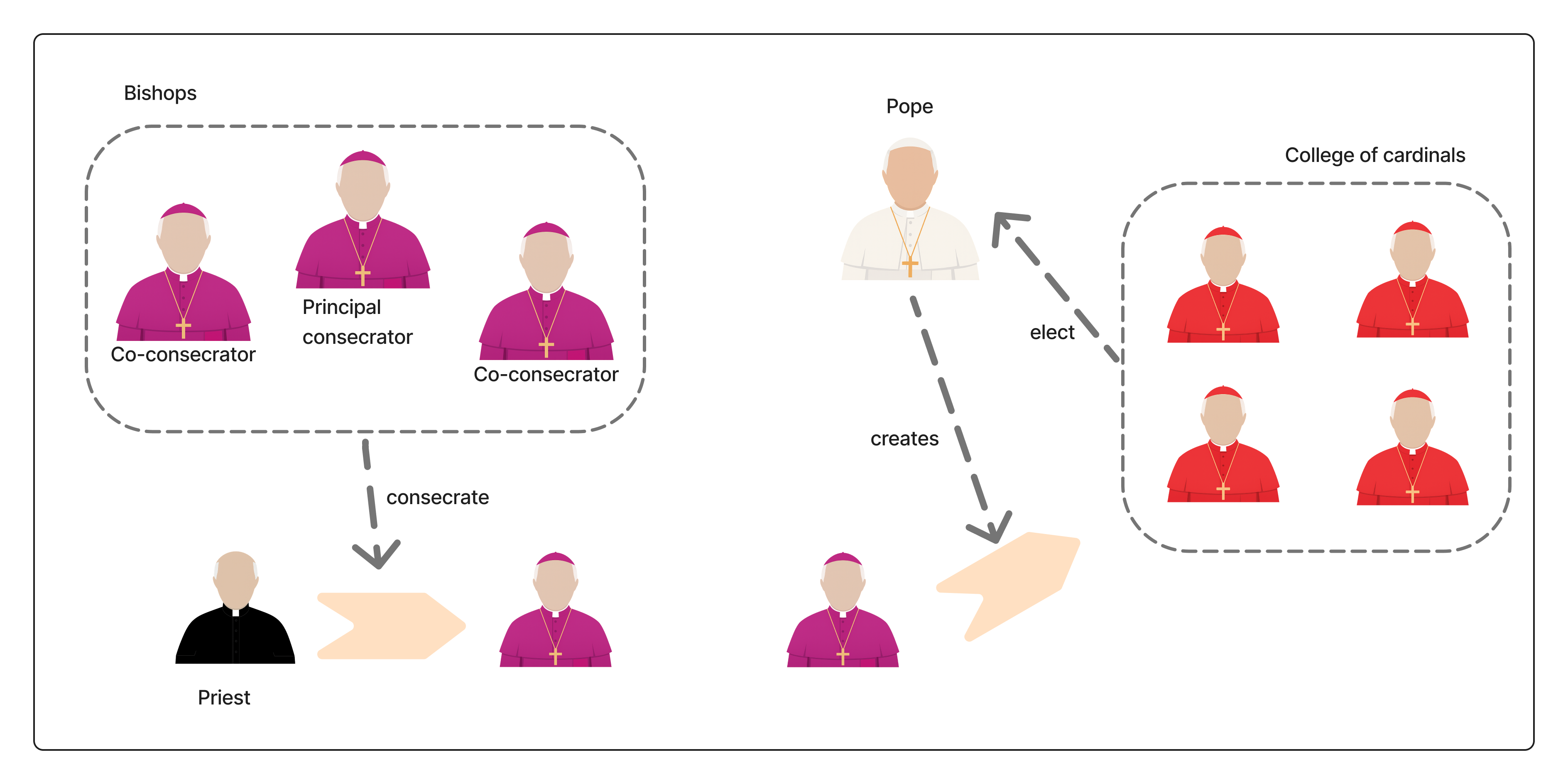}
		\caption{\textbf{Visual description of the hierarchical structure of the episcopate}: Left panel depicts the consecration process requiring three bishops (a principal consecrator and two co-consecrators) to maintain apostolic succession through ecclesiastical genealogy. Right panel illustrates the appointment of cardinals by the Pope and their role as papal advisors and in papal elections.}
		\label{fig:infograph}
	\end{figure}

	% Introduction/Background to our problem
	The Roman Catholic Church constitutes one of the most enduring religious and political institutions in history, with a continuous presence in both time and space, extending over two millennia on a global scale. To understand the size of this institution, it is worth noting that it is possible to track records of at least 34,573 bishops in the history of the Church, according to the website Catholic-Hierarchy.org~\cite{catholic-hierarchy}. 
	Over this extensive period, the Church has developed a sophisticated hierarchical structure aimed at administering its global constituency. At the foundation of this ecclesiastical organization lies the episcopate: each bishop is entrusted with both spiritual leadership and administrative authority. While most bishops are responsible for the governance of a diocese—an ecclesiastical jurisdiction corresponding to a defined territorial unit—some serve in non-territorial roles, including administrative, diplomatic, or curial positions. The ordination of a new bishop is traditionally performed by at least three consecrating bishops, one of which is necessarily the Principal Consecrator (PC) and the other two the Co-Consecrators (CC) (see Figure~\ref{fig:infograph}). This process establishes a form of ecclesiastical genealogy, which is believed to reflect Apostolic Succession—that is, an unbroken line of episcopal consecrations tracing back to the Apostles.
	
	Within this hierarchical framework, certain bishops may be appointed to specific offices, including that of cardinal. A cardinal does not receive a distinct sacramental or ecclesiastical status, as he remains within the same apostolic ministry conferred at his episcopal ordination. However, the appointment to the cardinalate entails a significant increase in authority and responsibility: cardinals serve as principal collaborators of the Pope in the governance of the universal Church and, provided they are under the age of 80, are eligible to participate in the conclave for the election of a new pontiff. Moreover, due to their role, cardinals can be considered probable candidates to exert a significant influence on the diffusion of opinion within the Church. 
	In recent years, the intricate structure of the Roman Catholic Church has begun to be studied using network science techniques, both in relation to social dynamics~\cite{bullivant2022power} and for health purposes~\cite{negron2014leadership}.
	
	% Our problem/data: network and textual data
	In this study, the doctrinal influence of hierarchical structures within the Church is investigated. We model the relationships connecting bishops through lines of episcopal consecration as a directed network. In this graph each node represents a bishop and a directed edge from one node to another exists if, and only if, the former participated in the consecration of the latter.
	
	% Already existing techniques:
	% Network 
	The resulting consecration network is a DAG (Directed Acyclic Graph), which is a usual way to represent and analyze genealogical trees \cite{dag_gen1}. To investigate the structural relationships among cardinals, especially by establishing pairwise connections, is particularly useful to analyze specific network motifs generated by the consecration process. This strategy is a well-established approach in complex network analysis~\cite{milo2002network}. The search for motifs has been extensively used for various purposes, from biological networks analysis \cite{wong2012biological}, to identification of strong bonds in social networks \cite{rotabi2017detecting}.

	% Topic modelling
	To analyze the most relevant topics of interest to bishops, as well as their perspectives on these issues, various topic modeling techniques can be employed. These models, which belong to the broader class of probabilistic models for text, are designed to uncover the latent thematic structure within large corpora \cite{topic_modelling_latent}. 
	One of the most popular methods is Latent Dirichlet Allocation (LDA), a hierarchical Bayesian model organized across three levels: words, topics, and documents \cite{blei2003latent}. In the context of Catholic discourse, for example, LDA has been used to identify recurring themes across a set of 37 homilies delivered by the priest that founded "Opus Dei" \cite{Tagliapietra02012025}. Another model that recently succeeded in identifying clusters of topics in textual data is the transformer based model BERTopic \cite{grootendorst2022bertopic}. BERTopic supports a wide range of topic modeling techniques and clustering algorithms, also allowing to have an indefinite number of possible clusters, letting the themes naturally emerge from the data. The possibility to decline BERTopic with different settings suitable for our purpose led us to choose this technique instead of others. 
	
	% Our contriubtion
	In particular, this study identifies ten salient topics emerging from textual summaries and corpora of statements made by living cardinals throughout their lifetimes. These include debates on LGBTQIA+ individuals within the Church, the role of women in ecclesiastical contexts, and broader issues such as bioethics, migrations and liturgy. Each topic is then associated with individual cardinals, allowing for the extraction of their respective polarity on each issue. Subsequently, we compute the ideological distance for each pair of cardinals, for each topic. 
	Finally, we estimate both univariate and multivariate linear regression models where the dependent variable corresponds to the ideological distances, while the independent variables include both the relationships identified through network motifs and biographical metadata associated with each cardinal. According to the parameters of the linear regression, we infer if there is a correlation between the opinions and the relationships.

	\section{Methods}
	\label{sec:method}
	
	\subsection{Data description}
	
	%Bishops are the the highest rank of ordained clergy in the Catholic Church, coordinate its ministry and manage its administration.
	%Bishops are ordained through the sacrament of consecration, performed by one other bishop (the {\itshape principal consecrator}), assisted by, typically, two other bishops ({\itshape co-consecrators}).
	%This creates a genealogy that tradition and doctrine trace back to the apostles of Jesus Christ, first bishops of the Church. 
	%The pope, bishop of Rome and head of the Church 
	
	The source of the genealogical data is the database {\itshape Catholic Hierarchy}~\cite{catholic-hierarchy}, reporting biographical data on living and deceased bishops, as well as their genealogy, and whether they were elevated to cardinal. We identified 250 living cardinals and discarded 5 who were never consecrated bishops: a possible, albeit rare, occurrence.
	For each of them, we retrieved their principal consecrators (PC), co-consecrators (CC) as well as the bishops that they consecrated in either role. We also retrieved their date of birth and country of origin. The source of the opinion data is the database {\itshape The College of Cardinals Report}~\cite{college_of_cardinals} that contains textual information about living cardinals.
	Both datasets are publicly accessible at the URLs provided.
	
	\subsection{Genealogical relationships}
	
	The episcopal lineage can be naturally represented as a Directed Acyclic Graph (DAG), where each node corresponds to a bishop, and a directed edge from bishop $i$ to bishop $j$ indicates that $i$ participated in the consecration of $j$.
	Specifically, causality ensures that the graph is acyclic but the tree-like structure is likely to be broken as one bishop can consecrate a new bishop together with the bishop who consecrated him.
	%Due to the chronological and hierarchical nature of consecration, the graph is acyclic: no bishop can be part of a cycle or consecrate someone who, directly or indirectly, consecrated him. 
	According to canonical tradition, each new bishop is be consecrated by at least three existing bishops, one principal and two co-consecrators: The in-degree is thus most often equal to $3$.
	
\begin{figure}[h]
		\centering
		\includegraphics[width=0.9
		\textwidth]{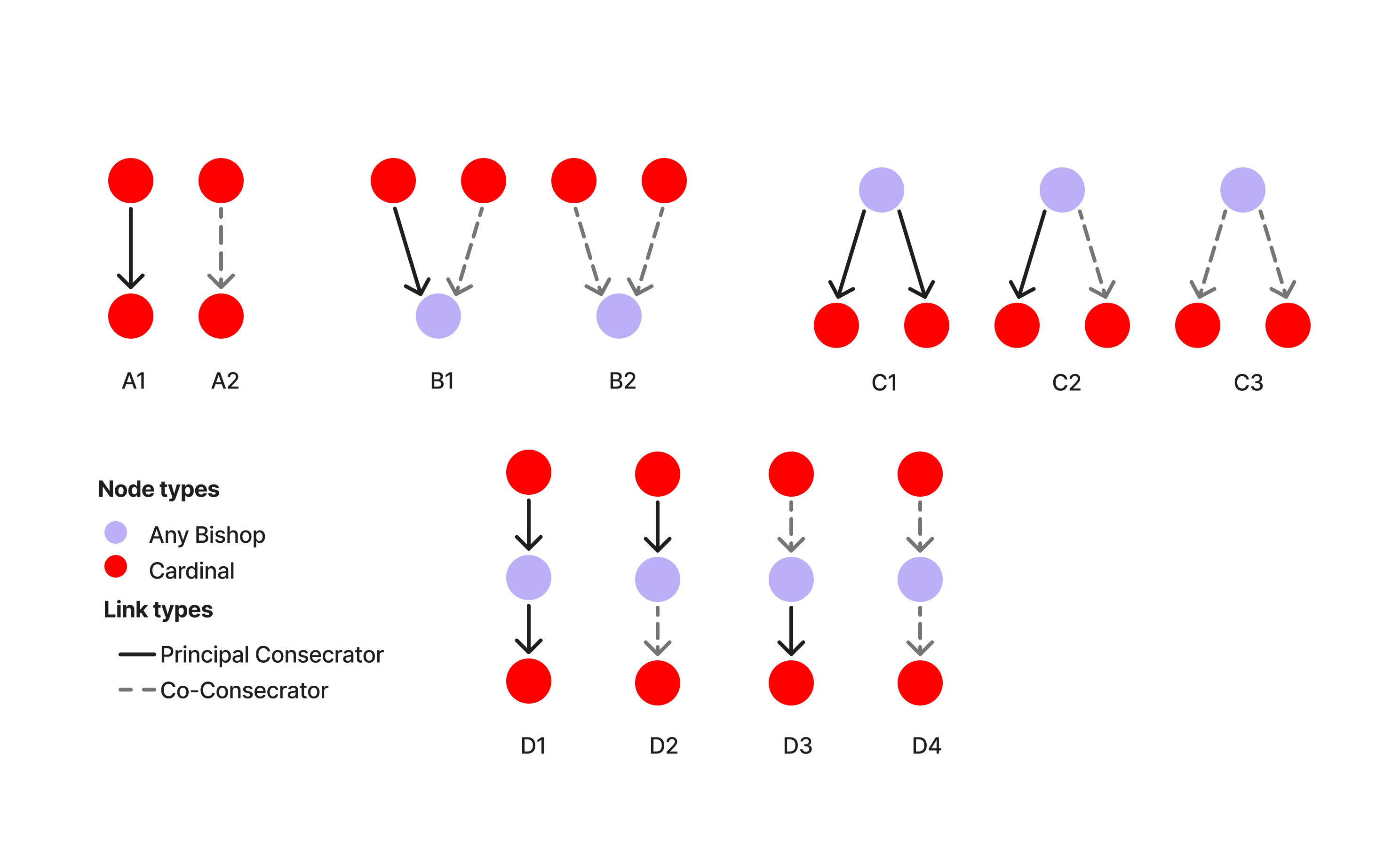}
		\caption{\textbf{Motif structure.} Schematic illustration of the motifs of interest. The possible motifs involving a pair of Cardinals (red) in the genealogy network can be split in four classes $A,B, C,D$. Motifs can also involve Bishop nodes (violet) who may not be Cardinals but participate in the motif. We can further distinguish this classification according to whether the Cardinal had the role of PC (continuous link) or CC (dashed link).}
		\label{fig:motif}
	\end{figure}

	In this genealogy, we identified four sets of motifs that we posit might be associated with opinion alignment. Specifically, we focused on the following pairwise genealogical relations among the living cardinals. Each motif is further characterized by the role of the consecrating bishop(s) --~either a PC or CC~-- allowing for a finer-grained analysis of structural roles. We refer to the motifs involving the type of consecration link as micro-motifs. Figure~\ref{fig:motif} illustrates the possible link combinations for each genealogical micro-motif listed here:
	
	\begin{itemize}
		\item (A) \textbf{Consecrator–Consecrated}: a cardinal consecrates a bishop who later becomes a cardinal:
		\begin{itemize}
			\item (A1) the consecrator is the principal consecrator;
			\item (A2) the consecrator is one of the co-consecrators;
		\end{itemize}
		\item (B) \textbf{Joint-Consecrators}: two cardinals jointly consecrate the same bishop:
		\begin{itemize}
			\item (B1) one of the consecrators is the principal consecrator;
			\item (B2) both consecrators are co-consecrators;
		\end{itemize}
		\item (C) \textbf{Joint-Consecrated}: a bishop consecrates two bishops who later become cardinals:
		\begin{itemize}
			\item (C1) the consecrator is the principal consecrator of both consecrated bishops;
			\item (C2) the consecrator is the principal consecrator for only one of the two consecrated bishops;
			\item (C3) the consecrator is a co-consecrator of both consecrated bishops;
		\end{itemize}
		\item (D) \textbf{Bridged Consecrator–Consecrated}: a cardinal consecrates a bishop (cardinal or not), who in turn consecrates a bishop who later becomes a cardinal:
		\begin{itemize}
			\item (D1) both bishops are principal consecrators;
			\item (D2) only the first bishop is a principal consecrator;
			\item (D3) only the second bishop is a principal consecrator;
			\item (D4) both bishops are co-consecrators.    
		\end{itemize}
	\end{itemize}
	
	It should be noted that these motifs are not mutually exclusive and may overlap in the network structure. For example, a pair of cardinals involved in multiple consecrations may simultaneously participate in several motif types, or motifs may combine to form larger configurations such as triangles or chains of interactions.
	%However, in this study, we do not explicitly account for such higher-order overlaps, focusing instead on the enumeration and analysis of individual first-order motifs.
	
	%\end{center}

	\subsection{Topic identification}
	
	%Pipeline for topic modeling : cleaning text, removing stopwords, lemming,stemming, tokenizing. Removing too common words.
	
	The textual dataset is built using the life summary of living cardinals. We have the textual data connected to the corresponding person.
	We applied the \texttt{en-core-web-sm} model from the \texttt{spaCy} library~\cite{spacy2020} to the textual data of each cardinal, %\footnote{\url{https://spacy.io/models/en\#en_core_web_sm}} 
	to divide paragraphs into meaningful chunks. We then embedded each chunk using \texttt{all-mpnet-base-v2}, a sentence-transformers model of HuggingFace~\cite{huggingface}.
	%\footnote{\url{http://huggingface.co/sentence-transformers/all-mpnet-base-v2}}.
	We then applied BERTopic \cite{grootendorst2022bertopic} for topic extraction from the embeddings, with the HDBSCAN clustering algorithm.
	This approach enabled the model to identify topics emerging directly from the data, without imposing prior constraints. At the same time, we guided BERTopic by providing predefined topic labels and representative keywords, which are listed in Appendix \ref{appendix:keywords}. \\
	The labels reflected themes we expected to encounter, while the keywords helped to more precisely characterize each topic. Following topic extraction, we manually reviewed all identified topics and merged those that were semantically similar but had been split into separate clusters by the algorithm.
	
	\subsubsection{Opinion identification}
	We used the Large Language Model (LLM) \texttt{gemini-2.5-flash-preview-05-20}~\cite{gemini2025} to determine the ideological orientation of the cardinals across the different topics. LLMs have demonstrated strong performance in evaluating the quality and content of texts~\cite{llm-evaluation}, making them well-suited for systematic ideological classification.\\
	For each topic, all sentences associated with a given cardinal were provided as input to the LLM, which was then prompted to assess the cardinal’s stance on that topic. The model classified each position into one of five categories: very conservative, conservative, neutral, progressive, or very progressive.
	
	To ensure reliable and accurate assessments, we implemented several structured prompt engineering techniques \cite{google_PromptEng, systematic_promptEng}. Our approach involved clearly defining the LLM's role, establishing explicit contextual frameworks for the evaluation, and providing explicit indicators and criteria to help it distinguish between progressive and conservative viewpoints. The complete prompt specification can be found in the Appendix ~\ref{appendix:prompt} .

	\color{black}
	
	\subsection{Statistical models}
	
	To investigate the factors influencing ideological distances between cardinals, we perform both univariate and multivariate linear regression analyses using a range of predictor variables, including cardinal metadata and genealogical motifs. Specifically, we consider the following features for each pair of cardinals:
	the absolute age difference, a binary indicator for whether the cardinals are from the same country, and the presence of macro- and micro-motifs in their genealogical relationships.
	
	For each predictor variable $X$, we fit a simple linear regression model accross all cardinal pairs $(i,j)$ as in
	\begin{equation}
		Y_{(ij)}^\alpha = q + \beta X_{(ij)} + \varepsilon,
	\end{equation}
	where $Y_{(ij)}^{\alpha}$ represents the ideological distance between a cardinal pair $(i,j)$ on a given topic $\alpha$. $q$ is the intercept, $\beta$ is the slope coefficient representing the association between the predictor and the ideological distance, and $\varepsilon$ is the error term. 
	To examine the joint effects of multiple predictors while controlling for potential confounding variables, we also fit a multiple linear regression model including different predictor variables simultaneously
	\begin{equation}
		Y_{(ij)}^\alpha = q + \sum_\alpha \beta^\alpha X_{(ij)}^\alpha + \varepsilon,
	\end{equation}
	where $X^\alpha_{(ij)}$ represents the predictor variable $\alpha$, and $\beta^\alpha$ is the corresponding regression coefficients representing the independent contribution of each predictor to the ideological distance.
	Moreover, we assess potential multi-collinearity among predictor variables by computing the Variance Inflation Factor (VIF) for each covariate \cite{obrien2007caution}.

	%Given the heterogeneous nature of our dataset, we also implemented a balanced sampling approach to address potential biases arising from unequal representation of features combinations that might confound our analysis. For each categorical predictor (e.g., macro motif A), we performed the following stratified sampling procedure:
	
	%\begin{enumerate}
	%    \item \textbf{Stratification}: Divided the data into homogeneous groups by fixing combinations of other relevant features (e.g., same region, different hierarchical order, same undirected network distance).
	
	%    \item \textbf{Group-wise counting}: Within each group, counted the number of cardinal pairs exhibiting the categorical predictor.
	
	%    \item \textbf{Balanced resampling}: For each group, created a balanced dataset by:
	%    \begin{itemize}
		%        \item Including all ideological distance values associated with the focal feature
		%        \item Randomly sampling twice this number of ideological distance values from pairs \textit{not} associated with the focal feature within the same group
		%    \end{itemize}
	
	%    \item \textbf{Aggregation}: Combined all balanced groups to form the final balanced dataset for analysis.
	%\end{enumerate}
	
	%This approach ensures that each analysis is conducted on a dataset where the categorical predictor has a 1:2 ratio (presence:absence), while maintaining the natural distribution of other confounding variables within each group.

	To assess the statistical significance of the regression coefficients in both univariate and multivariate models, while accounting for the network structure of our data, we employ the Mantel permutation test \cite{krackhardt1988predicting}.  For each model, we compute empirical $p$-values by randomly permuting the ideological distance values across cardinal pairs 100 times. This permutation-based approach provides robust inference without relying on distributional assumptions and appropriately adjusts for the dependencies present in dyadic (network) data. Statistical significance is evaluated at a threshold of $p<0.05$ or $p<0.1$ .
	The analysis is first conducted using univariate regression models for each predictor variable within each specific topic. Variables exhibiting statistically significant associations (at $p<0.05$) are subsequently included in a multivariate regression model for that topic,  to which the Mantel permutation test is applied again ($p<0.1$). This procedure enables the identification of key features associated with the ideological positioning of cardinals on the topic under consideration.

	%The explanatory power and accuracy of our analysis were also assessed through the determination of two key performance metrics, the coefficient of determination ($R^2$), calculated as the proportion of variance in ideological distance explained by the predictor variable(s), and the mean absolute error ($MAE$), computed as the average absolute difference between observed and predicted ideological distances. 
	
	%These metrics allow for comparison of model performance both within and across univariate and multivariate specifications, providing insights into the relative importance of individual predictors and the added value of the comprehensive multivariate approach.

	\section{Results and Discussion}
	
	% -Do pair of cardinals in a consecrational relation (motifs) show higher ideological similarity than random pairs? 
	
	% -Do ideological similarity decreases with genealogical distance (e.g. father -> grandfather -> great-grandfather - easy test to do????)
	
	% -Do network positions correlate with actual Vatican policy outcomes?
	
	% -Alternatives: Geographic proximity, organization, etc.
	%

	\subsection{Opinion identification} 
	
	The themes identified with topic modeling are summarized in Table~\ref{tab:cardinals_topics}, including the corresponding number of cardinals who addressed each theme in our dataset.
	Each topic, labeled with a descriptive name, is briefly clarified in the following.
	\textit{LGBTQIA+} refers to discussions concerning the role of same-sex couples within the Church and broader issues related to gender theory.
	\textit{Bioethics} encompasses debates surrounding euthanasia, abortion, and contraception.
	The topic labelled \textit{Women} addresses the question of female ordination and the potential for expanding women's roles within the Church.
	\textit{Synodality} relates to ecclesial governance, specifically whether decision-making should be more shared among members of the clergy --~synodal~-- rather than centralized in the pope and the Roman curia.
	\textit{Migration} concerns the attitudes of the Church toward migrants and displaced populations.
	\textit{Climate} refers to the opinions pertaining to climate change and the preservation of the environment.
	The topic of \textit{Marriage} involves the Church’s position on the inclusion of divorced and civilly remarried individuals in its sacramental and communal life. \textit{Celibacy} explores the possibility of allowing priests to marry and have families.
	\textit{COVID-19} refers to perspectives on public health measures adopted during the COVID-19 pandemic, such as vaccination, mask mandates, lockdowns and movement restrictions.
	Finally, \textit{Liturgy} focuses on contemporary attitudes toward the celebration of the Latin Mass and its role in the life of the Church today.

	\begin{table}[htbp]
		\centering
		\caption{{\textbf{Number of cardinals addressing each topic}. For each theme, the table reports the number of cardinals who were found to have addressed the topic in our dataset.}}
		\label{tab:cardinals_topics}
		\begin{tabular}{cc}
			\toprule
			Topic & Number of cardinals \\
			\midrule
			\textit{Bioethics} & 67 \\
			\textit{Celibacy} & 28 \\
			\textit{Climate} & 43 \\
			\textit{COVID-19} & 33 \\
			\textit{LGBTQIA+} & 92 \\
			\textit{Liturgy} & 49 \\
			\textit{Marriage} & 52 \\
			\textit{Migration} & 54 \\
			\textit{Synodality} & 81 \\
			\textit{Women} & 53 \\
			\hline
		\end{tabular}
	\end{table}
	
\begin{figure}
		\centering
		\includegraphics[width=\textwidth]{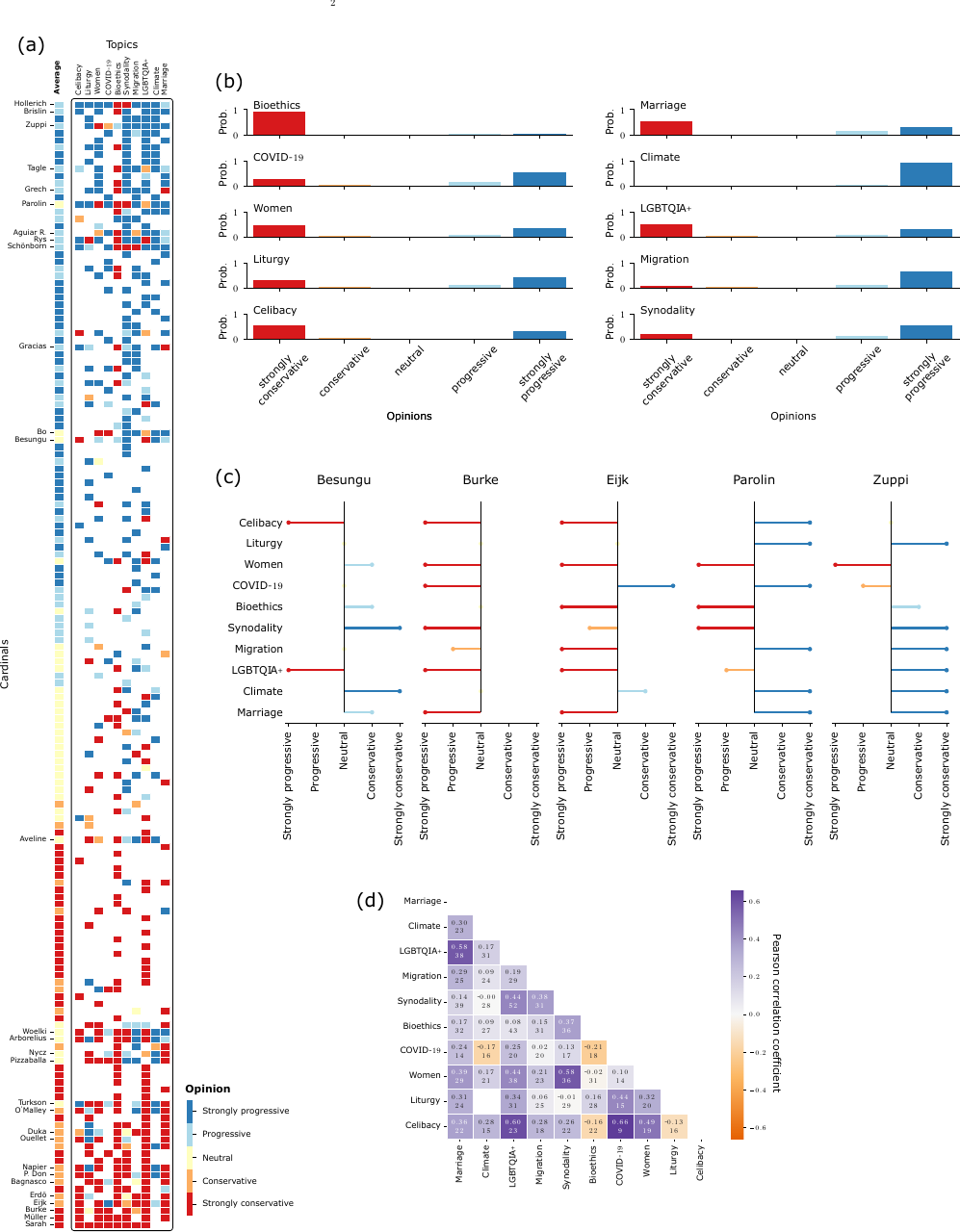}
		\caption{\textbf{Ideological spectrum of the cardinals.} \textbf{(a)} Heatmap with the opinion score assigned to Cardinals across the key topics identified from the textual data. Cell colors range from dark blue (strongly progressive) to dark red (strongly conservative). White cells indicate topics for which it was not possible to determine the opinion. Cardinals are sorted by the overall sum of their scores, while the average across all topics is displayed in the leftmost column. %Names are shown only for those with scores on at least seven topics.
			\textbf{(b)} Topic polarization as given by the normalized counts of the scores.
			%For bioethics and climate we cannot see a strong division between the two ideology. For other topics, for example the role of women in the church structure
			\textbf{(c)} Individual opinions of some selected Cardinals. \textbf{(d)} Pearson correlation matrix between topics based on Cardinals' ideological positions.}
		\label{fig:ideology}
	\end{figure}

	To visualize the opinion of each cardinal across various topics, we present a heatmap in Figure\ref{fig:ideology}(a), where rows represent individual cardinals and columns correspond to topics. Rows display only the names of cardinals who have identifiable opinions on at least seven topics.
	Additionally, the mean score of each cardinal is computed and shown in the designated leftmost column.
	A white cell indicates that the opinion could not be determined, either because the cardinal did not publicly expressed a view on the topic or because our data did not contain it.
	Otherwise, the cell is colored according to the assigned score: dark red for very conservative views, red for conservative, light yellow for neutral, light blue for progressive, and dark blue for very progressive positions.
	On several topics a clear polarization emerges, with many cardinals adopting either very conservative or very progressive positions. \\
	This pattern is also evident in Figure\ref{fig:ideology}(b), which shows a bar plot of the frequency distribution of opinion scores across themes. Most topics display a strongly bimodal pattern, reflecting sharp ideological divides. Notable exceptions are bioethics, climate, and migration: in these cases, the majority of cardinals express similar views: strongly conservative on bioethics, strongly progressive on climate, and progressive on migration.\\
	In Figure\ref{fig:ideology}(c), we highlight a selection of representative cardinals to illustrate their positions on the various topics, where available. The selected cardinals differ in their overall degree of polarization, ranging from consistently strong ideological stances to more nuanced or topic-dependent views.

	Finally, we measured the correlation of opinion polarization across topics (see Figure~\ref{fig:ideology}(d)). For any two topics, we computed the Pearson correlation coefficient between the cardinal scores , including only the cardinals who had an identifiable opinion on those topics.
	A strong positive correlation is observed between topics such as the openness towards \textit{LGBTQIA+} issues and the inclination to allow a full participation in the sacraments, including Holy Communion, for the divorced and civilly remarried couples.  Cardinals who hold conservative views on \textit{LGBTQIA+}matters also tend to adopt conservative positions on related themes such as marriage and priestly celibacy. In contrast, certain topics, such as migration and liturgy, exhibit negligible or no correlation.
	This outcome suggests that cardinals adopt consistently progressive or conservative views on matters pertaining to gender, family and sexuality.
	Beyond that, however, a univocal classification of a cardinal as progressive or conservative does not seem to emerge.
	Finally, correlations with the topics bioethics should be interpreted with care as nearly all cardinals were identified as strongly conservative on this issue,  reflecting a uniform stance of the Church.
	%In contrast, discourse surrounding COVID-19 tended to be more personal, with the majority of cardinals adopting progressive positions—advocating for the use of vaccines and masks. This divergence in tone and stance helps explain the observed negative correlation between these two topics.
	
	%mean cardinal Vanilla distance: 4.87   standard deviation 1.22
	%mean null-model Vanilla distance: 5.96   standard deviation 1.31

	\begin{table}[htbp]
		\begin{center}
			\caption{\textbf{Occurrences of the motifs}. For each category of motif, the table reports the number of times it was found in the genealogy network.}
			\begin{tabular}{*{2}{c}*{2}{c}*{3}{c}*{4}{c}} 
				\toprule
				\multicolumn{2}{c}{\textbf{A}} & 
				\multicolumn{2}{c}{\textbf{B}} & 
				\multicolumn{3}{c}{\textbf{C}} & 
				\multicolumn{4}{c}{\textbf{D}} \\
				\multicolumn{2}{c}{98} & 
				\multicolumn{2}{c}{467} & 
				\multicolumn{3}{c}{643} & 
				\multicolumn{4}{c}{98} \\
				\midrule
				\textbf{A1} & \textbf{A2} & \textbf{B1} & \textbf{B2} & \textbf{C1} & \textbf{C2} & \textbf{C3} & \textbf{D1} & \textbf{D2} & \textbf{D3} & \textbf{D4} \\
				56 & 42 & 297 & 165 & 369 & 106 & 168 & 6 & 27 & 20 & 45 \\
				\hline
			\end{tabular}
		\end{center}
		
		\label{tab:count_motif}
	\end{table}

	\subsection{Genealogy proximity and similarity of opinion}
	
	The statistical model detected the existence and significance of an association between the fact that two cardinals were genealogically linked by means of the selected motifs, and the distance between their views on a specific issue, measured as the absolute difference in their opinion score.
	We performed an independent analysis on each topic and included all possible pairs of cardinals who had a position on that topic. Then, 
	we estimated the coefficient of the linear relationship between the binary variable of the pair belonging to a given motif and their opinion distance.
	%Univariate and multivariate linear analyses potentially reveal a direct relationship between ecclesiastical genealogy and the ideological positions of cardinals.
	This means that a negative coefficient indicates that being part of a given motif is associated with smaller ideological distance, i.e., the cardinals having similar views on the topic.
	In the analysis we also included additional features who could potentially confound the relationship between genealogy and opinion, such as age difference and sharing the same country of origin.
	
	We first fit a univariate model, whose results are presented in Table~\ref{tab:univariate} from the Appendix~\ref{appendix:D}.
	Then, we included the variables exhibiting significant associations into multivariate models (Table~\ref{tab:multivariate}).
	We computed VIF and detected no evidence of multicollinearity.
	%Then we included the variables that were significant in the univariate models into multiva
	%The results of the multivariate model are presented in Table~\ref{tab:multivariate}, while those of the univariate model are shown in Table \ref{tab:univariate} of Appendix \ref{appendix:D}. 
	In both analyses, we exclude the topics of bioethics and climate due to their high polarization toward a single viewpoint, as illustrated in Figure \ref{fig:ideology}(b).
	
\begin{figure}[t]
		\centering
		\includegraphics[width=0.7
		\textwidth]{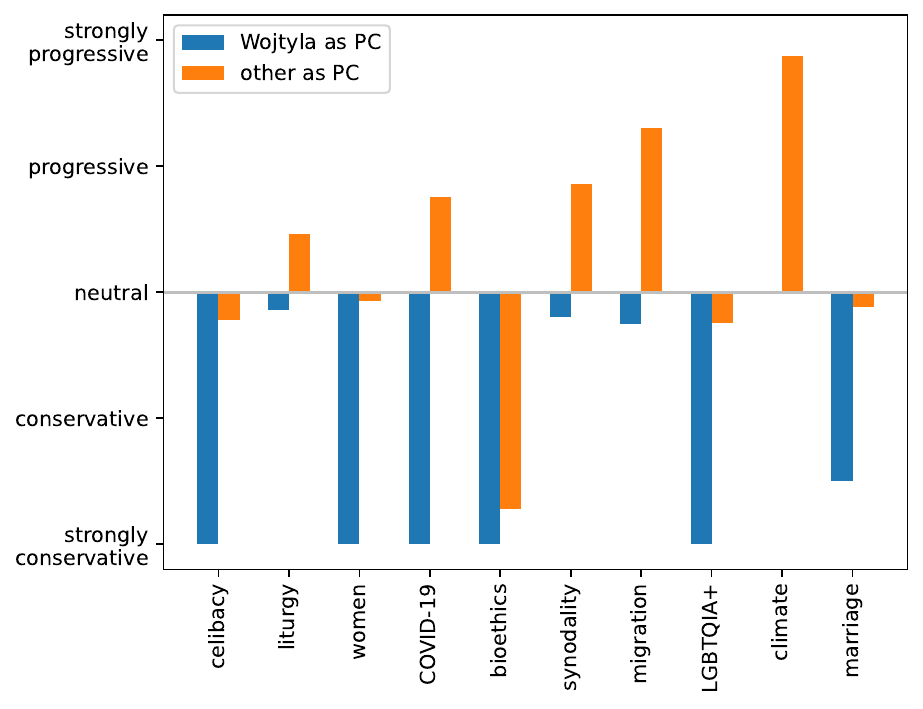}
		\caption{\textbf{Effect of motif C1 and role of John Paul II.} The figure reports the average ideological position across topics for cardinals involved in the C1 motif. The opinion of cardinals with John Paul II (Karol Wojtyła) as PC is in blue, the opinion of those with any other PC is in orange.}
		\label{fig:wojtyla}
	\end{figure}

	Across the different topics, no more than five features were significantly associated with ideological positioning in the univariate models.
	Expectedly, similarity in age is broadly associated with greater ideological similarity: generational differences may impact cardinal views in both moral and doctrinal issues.
	Notably, the C1 motif (see Figure~\ref{fig:motif}) is also consistently associated with similar views on topics pertaining to gender, family and sexuality. 
	%This suggests that cardinals who share the same principal consecrator are more likely to share similar opinions on 
	%The negative association of the C1-type motif suggests that when a bishop (cardinal or not) serves as the Principal Consecrator of two cardinals, those cardinals tend to hold more similar ideological positions.
	This suggests that consecration may be a driver, or at least a clustering factor, of opinion formation or transmission within the Church clergy.
	To further investigate this we identified the cardinals engaging in any C1 motif for which an opinion difference could be computed, and found that many (15 out of 31) had Karol Wojtyła (pope John Paul II) as their principal consecrator. We then compared the average opinion polarization of those 15 and the remaining 16 whose principal consecrator is not Wojtyła (see Figure~\ref{fig:wojtyla}). The episcopal offspring of John Paul II was consistently, and substantially, more conservative across the topics pertaining to gender, family and sexuality, and moderately more conservative on the other topics.
	This confirms the role of C1 as a predictor of ideological similarity as well as the role of the long papacy of John Paul II in shaping the ruling class of the Catholic Church.
	%These two features, identified as significant in the univariate model, remain consistently relevant in the multivariate model across different topics. However, their implications may vary depending on the specific context of each topic.
	
	%In the case of C1-type motifs, correlations are particularly relevant for the topics of Celibacy, LGBTQIA+, Marriage, and Women. For example, Figure \ref{fig:wojtyla}  shows that cardinals consecrated by John Paul II  (\color{red} Wojtyła \color{black}) $-$ the biggest C1-type consecrators and who is generally characterized as conservative $-$ also tend to hold more conservative views on all of these topics compared to the average ideological stance of other cardinals. 
	
	Beyond C1 motif, C3 was also significantly associated with opinion similarity in \textit{LGBTQIA+ i} in the univariate analysis, suggesting a possible effect of sharing a co-consecrator. This effect disappeared, however, after adjusting for C1. This suggests that C1 acts as a confounder: sharing the principal consecrator is associated with both sharing a co-consecrator and being ideologically similar, but sharing a co-consecrator is not {\itshape per se} associated with greater ideological similarity.
	%note that the A particularly interesting aspect of our multivariate analysis is that C3-type motif (Co-Consecrator relationships) loses statistical significance when the C1 motif is included in the model for the LGBTQIA+. 
	%This phenomenon occurs because C1 acts as a confounder for C3, meaning that the apparent effect of C3 on ideological similarity is actually mediated through the underlying C1 relationships. 
	%This disappearance of C3 effects when controlling for C1 actually suggests that any apparent ideological similarity between cardinals sharing CCs may be spurious, creating indirect connections that do not represent independent influence channels.
	Therefore, ideological proximity seems to be mediated primarily by the influence of the principal consecrator and not the co-consecrator. 
	%Should this be dropped?

	%The opinions on this topics of the most PCs consecrating cardinals is John Paul II (Wojtila), who is one of the most consecrator

	The analysis of ideological positions on \textit{Celibacy} is also interesting and differs from the \textit{LGBTQIA+} topic. Here we observe contrasting effects between different genealogical motifs: C1 is still associated with ideological proximity, but C2 is significantly associated with ideological distance, maintaining statistical significance in the multivariate mode.
	This suggests that the selection of co-consecrators may follow different strategic considerations that actually promote ideological diversity.
	This indicates that these motifs capture independent aspects of ideological transmission and network structure specific to this topic. 
	
	The ideological analysis on \textit{Liturgy} reveals another genealogical effect beyond C1, as motif A (direct consecration relationship) is also associated with ideological proximity. 
	The prominence of these motifs in the liturgical topic reinforces the hypothesis that direct genealogical transmission may serve as a primary mechanism for ideological propagation in more than one topic.
	
	Finally, sharing the same country of origin is associated with having a similar opinion on \textit{Migration}, consistent with the topic being of high geopolitical relevance.

	\begin{table}[htbp]
		\centering
		\caption{\textbf{Results of the multivariate linear regression model}. For each topic and feature, we here present the coefficients with its associated $p$-values between parenthesis. Those values with $p<0.1$ are bold highlighted.}
		\label{tab:regression_results_part1}
		\begin{tabular}{cccccccc}
			\toprule
			Topic & A & B2&  C1 & C2 & C3 & Year of birth & Country \\
			\midrule
			\textit{Celibacy} &  - & -& \textbf{-1.95  ( < 0.01)} & \textbf{2.05 (0.03)} & - & - & -\\
			\textit{LGBTQIA+} & - & - & \textbf{-1.76  ( < 0.01)} & - & 0.30 (0.46) & - & - \\
			\textit{Liturgy} & -1.9 (0.1) &  - & \textbf{-1.7 (0.02)} &  - & - & - \\
			\textit{Marriage} & - & - & \textbf{-0.98 (0.03)} & - & - & \textbf{0.01 (0.09)} & - \\
			\textit{Migration} & - &- & - & - & - & \textbf{-0.02 (<0.01)} & \textbf{-0.45 (<0.01)} \\
			\textit{Sinodality} & - & \textbf{0.47 (<0.01)} & - & - & - & \textbf{0.02 (0.03)} & - \\
			\textit{Women} & - & - & \textbf{-1.00 (0.07)} & - & - & \textbf{0.02 (0.01)} & - \\
			\hline
		\end{tabular}
		
		\label{tab:multivariate}
	\end{table}
	
	\section{Conclusion}
	
	This study presents limitations. First, the reliance on textual data to infer ideological orientation introduces a potential bias: the absence of a recorded statement may reflect either a deliberate silence by the cardinal or the absence of accessible data, and we could not distinguish between these two cases.
	Second, the textual dataset is limited to living cardinals; inclusion of non-cardinal bishops could reveal broader or distinct structural patterns of ideological alignment. Additionally, the exclusion of deceased figures, notably former popes, whose influence likely persists in the views of living cardinals they consecrated, thereby omitting potentially critical elements in the genealogical structure.
	If such data were available, it is plausible that motifs encoding direct relationships (A motifs) currently found to be less influential would gain in explanatory power.
	
	In sum, the analysis demonstrates that ecclesiastical genealogy exerts a measurable influence on the ideological configuration of the Catholic hierarchy. Motifs involving shared consecrators, particularly the principal consecrator, are consistently associated with ideological proximity across topics.
	These results provide quantitative support for the hypothesis that doctrinal and moral alignment is not solely an individual process but is structured through hierarchical transmission, suggesting that the mechanisms of episcopal consecration may play a role in the propagation of belief systems within the Church.

	\section{Acknowledgements}
	This work is the output of the Complexity72h workshop, held at the Universidad Carlos III de Madrid in Leganés, Spain, 23-27 June 2025, \url{https://www.complexity72h.com}.
	
	\newpage
	
	\appendix
	\section{Keywords and labels for topic modelling}
	\label{appendix:keywords}
	
	In this section, we provide the labels we used to force BERTopic to have certain topics and the keywords we passed to the model to find them.

	\noindent\fbox{%
		\begin{minipage}{\textwidth\fboxsep\fboxrule}
			\setlength{\parindent}{0pt}
			\setlength{\leftskip}{1em}
			\setlength{\rightskip}{1em}
			
			{\ttfamily
				{
					\textbf{"LGBT"}: "LGBT, same-sex couples, homosexual, blessing of gay couples, Fiducia Supplicans on same-sex blessings" \\
					\textbf{"poor"}: "poverty, migrants, refugees, solidarity, dignity, peripheries, fraternity, social justice, good government, inequality, exclusion, vulnerable, marginalization, hunger, charity, almsgiving, economic injustice, ntegral development, basic needs, human rights, common good, social doctrine, option for the poor" \\
					\textbf{"peace"}: "peace, war, justice, violence, disarmament, military, reconciliation, Fratelli tutti, international relations" \\
					\textbf{"ecumenism"}: "ecumenism, interreligious, interfaith, denominations, cultures, world religions, unity" \\
					\textbf{"environment"}: "climate-change, ecology, natural resources, respect for creation, environment, sustainability, common good, Laudato si', common home" \\
					\textbf{"synodality"}: "synodality, democratization, synod, consultation, collegiality, council, unity", \\
					"liturgy": "liturgy, rite, latin mass, novus ordo, vetus ordo, traditional mass", \\
					"divorced": "divorced, remarried, access to sacraments, irregular situations, Amoris Laetitia"
				}
			}
		\end{minipage}%
	}
	
	\newpage
	
	\section{Prompt LLM Evaluation Polarity}
	\label{appendix:prompt}
	In this section we provide the final prompt we used to perform the evaluation of the polarity of a certain text for a given topic.

	\noindent\fbox{%
		\begin{minipage}{\textwidth\fboxsep\fboxrule}
			\setlength{\parindent}{0pt}
			\setlength{\leftskip}{1em}
			\setlength{\rightskip}{1em}
			\textbf{Prompt:} 
			{\ttfamily
				"""
				You are an expert political analyst tasked with evaluating the political stance of Catholic cardinals' statements. 
				Your role is to objectively assess whether a given statement reflects a progressive or conservative position on a specific topic.
				
				**EVALUATION CRITERIA:** \\
				
				**Progressive Indicators:** \\
				
				- Emphasis on social change and reform \\
				- Support for expanded rights and inclusion \\
				- Calls for institutional change or modernization \\
				- Focus on social justice and equity \\
				- Openness to new interpretations or approaches \\
				- Emphasis on compassion over traditional doctrine \\
				
				**Conservative Indicators:** \\
				- Emphasis on traditional values and established doctrine \\
				- Resistance to change or reform \\
				- Upholding of institutional authority \\
				- Reference to historical precedent or scripture \\
				- Focus on moral order and traditional family structures \\
				- Caution about rapid social changes \\
				- Emphasis on doctrine over contemporary social movements \\
				
				**TASK:**
				Analyze the following statement by a cardinal on the topic of "{topic}":
				
				**Statement:** "{text}"
				
				**Topic:** {topic}
				
				**REQUIRED OUTPUT FORMAT:** \\
				Return me just a number as int or 'NaN'
				
				**EVALUATION GUIDELINES:** \\
				1. **Stance Classification:**
				- 2 stands for "strongly progressive": Clearly advocates for change, reform, or expanded inclusion \\
				- 1 stands for "mildly progressive": Advocates for change but not in a strong way.\\
				- -1 stands for "mildly conservative": Upholds traditional positions but not in a strong way \\
				- -2 is for "strongly conservative": Clearly upholds traditional positions or resists change in a strong way \\
				- 0 stands for "neutral": Balanced position or focuses on universal principles without clear political lean \\ 
				- "NaN": Ambiguous or insufficient information to make a determination because the text is not relevant to the topic \\

				3. **Focus on the specific topic:** Evaluate the statement specifically in relation to "{topic}", not general political orientation. \\
				4 **Not relevance to the topic:** If the text is not relevant to the given topic do not try to infere anything, just return 'NaN'.

				Provide your analysis now:"""}
		\end{minipage}%
	}
	
	\newpage
	
	\section{Network analysis}
	To evaluate the statistical significance of motif counts, we constructed a null model for comparison. We first identified the temporal range of living cardinals by examining their consecration years. The earliest living cardinal in our dataset---Cardinal Francis Arinze---was consecrated in 1965. We therefore defined all bishops consecrated from 1965 onward as \emph{living}. From this group of 9,068 living bishops, we randomly selected 245 individuals to serve as artificial cardinals in the null model.

\begin{figure}[t]
		\centering
		\includegraphics[width=\textwidth]{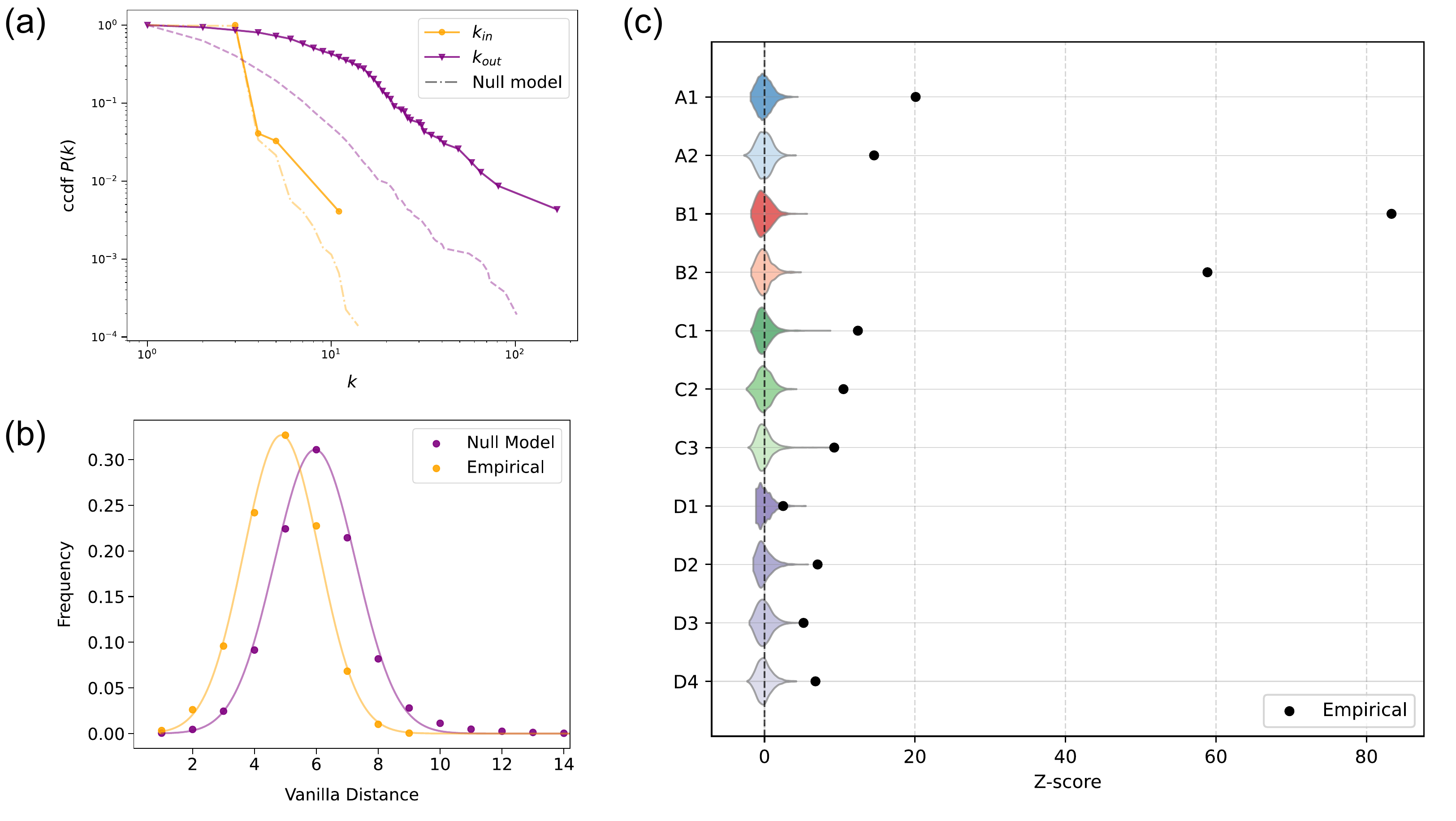}
		\caption{\textbf{Network structure and motif significance among living cardinal bishops.}
			\textbf{(A)} Out- and in-degree complementary cummulative distributions. The out-degree (purple triangles) is heavy-tailed, suggesting hubs; the in-degree (yellow circles) shows that most cardinals were consecrated by three bishops. Null model averages are shown as dashed and dash-dotted lines. 
			\textbf{(B)} Distribution of shortest-path distances among cardinal bishops. The empirical average distance is $4.87 \pm 1.22$, while null models average $5.96 \pm 1.31$, suggesting greater-than-expected cohesion. 
			\textbf{(C)} Violin plots for motif counts from 1,000 null-model realizations. Black dots represent $z$-scores for observed values, calculated as $z = \frac{x - \mu}{\sigma}$, indicating over-representation of specific motifs.}
		\label{fig:combined_panel}
	\end{figure}

	The out-degree complementary cummulative distribution (c.c.d.f.) presented in panel~(A) of Figure~\ref{fig:combined_panel} reveals the presence of hubs. The in-degree distribution shows that only a few cardinals have been consecrated by more than three bishops. On average, any two cardinals are separated by approximately five intermediary bishops, as shown in panel~(B). Remarkably, the distribution of shortest-path distances between cardinals closely follows a Gaussian profile. For comparison, the corresponding distribution computed from 100 randomized null-model networks also approximates a Gaussian but is centered around six. This systematic shift indicates that cardinals tend to be more closely connected to one another than would be expected by chance, suggesting a non-random, tightly-knit structure within the episcopal hierarchy. 
	
	We computed motif counts across 1,000 realizations of the reshuffled null-model network. Panel~(C) of Figure~\ref{fig:combined_panel} shows the resulting distributions of motif counts, along with the corresponding $z$-scores for the empirical observations. The $z$-scores are computed as $z = \frac{x - \mu}{\sigma}$, where $x$ is the observed motif count, $\mu$ is the average motif count across the null models, and $\sigma$ is the corresponding standard deviation. All motif counts deviate significantly from those produced by the null model, underscoring the non-random nature of cardinal selection and the structured configuration of priest-consecrator (PC) and co-consecrator (CC) links. This result confirms the structural specificity of the observed network.

	\section{Table univariate regression}
	\label{appendix:D}
	In this section, we report the parameters for the univariate lineal regression models for each predictor variable and topic.
	
	\begin{table}[htbp]
		\label{tab:univariate}
		\centering
		\caption{\textbf{Results of the univariate linear regression model}. For each topic and feature, we here present the coefficients with its associated $p$-values between parenthesis. Those values with $p<0.05$ are bold highlighted.}
		\label{tab:univariate}
		\begin{tabular}{ccccc}
			\toprule
			Topic & year of birth & country & order & region \\
			\midrule
			\textit{Bioethics} & \textbf{0.03 ( < 0.01)} & \textbf{0.29 (0.04)} & \textbf{-0.43 (<0.01)} & -0.04 (0.71) \\
			\textit{Celibacy} & 0.00 (0.90) & -0.08 (0.90) & 0.15 (0.45) & 0.08 (0.76) \\
			\textit{Climate} & \textbf{-0.01 ( <0.01)} & \textbf{0.23 (0.04)} & \textbf{0.17 (<0.01)} & -0.04 (0.60) \\
			\textit{COVID-19} & \textbf{0.02 (0.07)} & 0.22 (0.55) & -0.08 (0.64) & -0.04 (0.80) \\\
			\textit{LGBTQIA+} & 0.01 (0.14) & 0.28 (0.17) & -0.03 (0.65) & -0.12 (0.13) \\
			\textit{Liturgy} & 0.01 (0.30) & 0.09 (0.59) & 0.18 (0.11) & -0.22 (0.12) \\
			\textit{Marriage} & \textbf{0.01 (0.04)} & 0.03 (1.00) & 0.08 (0.52) & 0.05 (0.77) \\
			\textit{Migration} & \textbf{-0.02 (<0.01)} & \textbf{-0.51 (<0.01)} & \textbf{0.15 (0.06)} & -0.19 (0.13) \\
			\textit{Synodality} & \textbf{0.02 (<0.01)} & -0.17 (0.42) & -0.03 (0.54) & 0.10 (0.18) \\
			\textit{Women} & \textbf{0.02 (<0.01)} & -0.30 (0.24) & 0.09 (0.39) & 0.07 (0.61) \\
			\bottomrule
		\end{tabular}
	\end{table}
	
	\vspace{0mm}
	
	\begin{table}[htbp]
		\centering
		\begin{tabular}{cccc}
			\toprule
			Topic &A & A1 & A2 \\
			\midrule
			\textit{Bioetichs} &\textbf{0.79 (0.02)} & 0.32 (0.48) & \textbf{1.72 (<0.01)} \\
			\textit{Celibacy} &0.00 (1.00) & 0.00 (1.00) & 0.00 (1.00) \\
			\textit{Climate} &0.01 (0.82) & 0.10 (0.74) & -0.24 (0.81) \\
			\textit{COVID-19} &0.21 (0.94) & -1.79 (0.26) & 2.22 (0.18) \\
			\textit{LGBTQIA+} &-0.05 (0.91) & -1.41 (0.29) & 0.50 (0.58) \\
			\textit{Liturgy} &\textbf{-1.93 (0.08)} & -1.93 (0.14) & 0.00 (1.00) \\
			\textit{Marriage} &-0.45 (0.63) & -0.45 (0.76) & 0.00 (1.00) \\
			\textit{Migration} &-0.27 (0.56) & -0.10 (0.87) & -1.27 (0.39) \\
			\textit{Synodality} &-0.92 (0.17) & -0.66 (0.47) & -1.06 (0.17) \\
			\textit{Women} &-1.99 (0.12) & -1.99 (0.12) & 0.00 (1.00) \\
			\bottomrule
		\end{tabular}
	\end{table}
	
	\vspace{0mm}
	
	\begin{table}[htbp]
		\centering
		\begin{tabular}{lccc}
			\toprule
			Topics & B & B1 & B2 \\
			\midrule
			\textit{Bioethics}    & \textbf{0.25 (<0.01)}  & \textbf{0.30 (<0.01)}  & \textbf{0.43 (<0.01)}  \\
			\textit{Celibacy}     & 0.15 (0.38)  & 0.13 (0.33)  & 0.27 (0.61)  \\
			\textit{Climate}     & -0.08 (0.45) & -0.06 (0.73) & -0.09 (0.46) \\
			\textit{COVID-19}   & 0.22 (0.40)  & -0.29 (0.40) & \textbf{0.63 (0.10)} \\
			\textit{LGBTQIA+}    & 0.15 (0.21)  & 0.04 (0.79)  & 0.24 (0.24)  \\
			\textit{Liturgy}     & 0.13 (0.60)  & 0.39 (0.13)  & -0.58 (0.21) \\
			\textit{Marriage}    & 0.14 (0.59)  & 0.33 (0.43)  & 0.05 (0.76)  \\
			\textit{Mitigation}  & \textbf{-0.31 (0.06)} & -0.20 (0.58) & -0.37 (0.11) \\
			\textit{Synodality}  & 0.16 (0.22)  & -0.02 (0.77) & \textbf{0.46 (0.03)}  \\
			\textit{Women}      & -0.02 (0.94) & -0.50 (0.32) & 0.38 (0.33)  \\
			\bottomrule
		\end{tabular}
	\end{table}
	
	\vspace{0mm}
	\begin{table}[htbp]
		\centering
		\begin{tabular}{lcccc}
			\toprule
			Topics & C & C1 & C2 & C3 \\
			\midrule
			\textit{Bioethics}    & -0.07 (0.71) & -0.26 (0.48) & 0.06 (0.99)  & -0.06 (0.88) \\
			\textit{Celibacy}     & -0.40 (0.18) & \textbf{-1.96 (<0.01)} & \textbf{2.10 (0.04)}  & -0.33 (0.37) \\
			\textit{Climate}     & -0.24 (0.29) & -0.24 (0.53) & -0.24 (0.46) & -0.24 (0.75) \\
			\textit{COVID-19}   & -0.19 (0.74) & -0.79 (0.64) & -0.04 (0.85) & 0.00 (1.00)  \\
			\textit{LGBTQIA+}    & -0.59 (0.01) & -1.61 (0.00) & 0.33 (0.57)  & -0.71 (0.09) \\
			\textit{Liturgy}     & -0.04 (0.84) & 0.12 (0.76)  & -1.74 (0.01) & 0.33 (0.47)  \\
			\textit{Marriage}    & -0.26 (0.26) & -1.03 (0.03) & 0.45 (0.43)  & -0.16 (0.81) \\
			\textit{Migration} & -0.07 (0.82) & 0.49 (0.42)  & -1.07 (0.09) & -0.08 (0.89) \\
			\textit{Synodality}  & -0.07 (0.85) & 0.00 (0.91)  & -0.67 (0.22) & 0.08 (0.84)  \\
			\textit{Women}    & -0.37 (0.18) & -0.99 (0.07) & 0.68 (0.28)  & -0.90 (0.10) \\
			\bottomrule
		\end{tabular}
	\end{table}
	
	\vspace{0mm}
	
	\begin{table}[htbp]
		\centering
		
		\begin{tabular}{lccccc}
			\toprule
			Topics & D & D1 & D2 & D3 & D4 \\
			\midrule
			\textit{Bioethics}    & \textbf{0.68 (0.09)}  & -0.49 (0.69) & 0.00 (1.00)  & 0.51 (0.42)  & \textbf{0.95 (0.03)}  \\
			\textit{Celibacy}     & 0.76 (0.58)  & 2.09 (0.26)  & 0.00 (1.00)  & 0.00 (1.00)  & 0.09 (0.88)  \\
			\textit{Climate}     & -0.03 (0.95) & -0.24 (0.71) & -0.24 (0.74) & -0.24 (0.61) & 0.76 (0.21)  \\
			\textit{COVID-19}    & -0.88 (0.23) & 0.00 (1.00)  & -1.79 (0.30) & -1.29 (0.30) & 0.00 (1.00)  \\
			\textit{LGBTQIA+}   & -0.41 (0.59) & 2.10 (0.32)  & -0.24 (0.59) & -1.91 (0.31) & -1.91 (0.26) \\
			\textit{Liturgy}    & -1.43 (0.20) & 0.00 (1.00)  & -0.93 (0.53) & -1.93 (0.24) & 0.00 (1.00)  \\
			\textit{Marriage}    & 0.55 (0.72)  & 0.55 (0.67)  & 0.00 (1.00)  & 0.00 (1.00)  & 0.00 (1.00)  \\
			\textit{Migration}  & -0.56 (0.45) & 0.00 (1.00)  & -1.27 (0.44) & -0.77 (0.42) & 0.00 (1.00)  \\
			\textit{Synodality}  & 0.54 (0.63)  & 0.84 (0.48)  & -0.66 (0.68) & -0.66 (0.60) & 2.34 (0.24)  \\
			\textit{Women}      & 1.52 (0.28)  & 2.02 (0.30)  & 0.00 (1.00)  & 0.00 (1.00)  & 1.01 (0.59)  \\
			\bottomrule
		\end{tabular}
	\end{table}

	\newpage
	
	\bibliographystyle{unsrt}
	\bibliography{main.bib}

\begin{thebibliography}{10}

\bibitem{catholic-hierarchy}
{David M. Cheney}.
\newblock {Catholic Hierarchy}.
\newblock Accessed: 2025-06-24.

\bibitem{bullivant2022power}
Stephen Bullivant and Giovanni Radhitio~Putra Sadewo.
\newblock Power, preferment, and patronage: An exploratory study of catholic
  bishops and social networks.
\newblock {\em Religions}, 13(9):851, 2022.

\bibitem{negron2014leadership}
Rosalyn Negr{\'o}n, Bryan Leyva, Jennifer Allen, Hosffman Ospino, Laura Tom,
  and Sarah Rustan.
\newblock Leadership networks in catholic parishes: implications for
  implementation research in health.
\newblock {\em Social science \& medicine}, 122:53--62, 2014.

\bibitem{dag_gen1}
Imre Varga.
\newblock A case study of genealogical networks from network science
  perspective.
\newblock In E~Kusen and V~Chang, editors, {\em PROCEEDINGS OF THE 8TH
  INTERNATIONAL CONFERENCE ON COMPLEXITY, FUTURE INFORMATION SYSTEMS AND RISK,
  COMPLEXIS 2023}, COMPLEXIS, pages 47--52. INSTICC, 2023.
\newblock 8th International Conference on Complexity, Future Information
  Systems and Risk (COMPLEXIS), Prague, CZECH REPUBLIC, APR 22-23, 2023.

\bibitem{milo2002network}
Ron Milo, Shai Shen-Orr, Shalev Itzkovitz, Nadav Kashtan, Dmitri Chklovskii,
  and Uri Alon.
\newblock Network motifs: simple building blocks of complex networks.
\newblock {\em Science}, 298(5594):824--827, 2002.

\bibitem{wong2012biological}
Elisabeth Wong, Brittany Baur, Saad Quader, and Chun-Hsi Huang.
\newblock Biological network motif detection: principles and practice.
\newblock {\em Briefings in bioinformatics}, 13(2):202--215, 2012.

\bibitem{rotabi2017detecting}
Rahmtin Rotabi, Krishna Kamath, Jon Kleinberg, and Aneesh Sharma.
\newblock Detecting strong ties using network motifs.
\newblock In {\em Proceedings of the 26th international conference on world
  wide web companion}, pages 983--992, 2017.

\bibitem{topic_modelling_latent}
David~M. Blei.
\newblock Probabilistic topic models.
\newblock {\em Commun. ACM}, 55(4):77–84, April 2012.

\bibitem{blei2003latent}
DM~Blei, AY~Ng, and MI~Jordan.
\newblock Latent dirichlet allocation.
\newblock {\em Journal of Machine Learning Research}, 3, 2003.

\bibitem{Tagliapietra02012025}
Claudio Tagliapietra.
\newblock Automated text analysis in theology: An application.
\newblock {\em Theology and Science}, 23(1):55--71, 2025.

\bibitem{grootendorst2022bertopic}
Maarten Grootendorst.
\newblock Bertopic: Neural topic modeling with a class-based tf-idf procedure.
\newblock {\em arXiv preprint arXiv:2203.05794}, 2022.

\bibitem{college_of_cardinals}
{Pentin, E. and Montagna, D.}
\newblock {The College of Cardinals Report}.
\newblock Accessed: 2025-06-24.

\bibitem{spacy2020}
Matthew Honnibal, Ines Montani, Sofie~Van Landeghem, and Adriane Boyd.
\newblock spacy: Industrial-strength natural language processing in python.
\newblock 2020.

\bibitem{huggingface}
Thomas Wolf, Lysandre Debut, Victor Sanh, Julien Chaumond, Clement Delangue,
  Anthony Moi, Pierric Cistac, Tim Rault, Rémi Louf, Morgan Funtowicz, Joe
  Davison, Sam Shleifer, Patrick von Platen, Clara Ma, Yacine Jernite, Julien
  Plu, Canwen Xu, Teven~Le Scao, Sylvain Gugger, Mariama Drame, Quentin Lhoest,
  and Alexander~M. Rush.
\newblock Transformers: State-of-the-art natural language processing.
\newblock In {\em Proceedings of the 2020 Conference on Empirical Methods in
  Natural Language Processing: System Demonstrations}, pages 38--45, Online,
  October 2020. Association for Computational Linguistics.

\bibitem{gemini2025}
{Gemini Team, Google DeepMind}.
\newblock {Gemini 2.5: Pushing the Frontier with Advanced Reasoning,
  Multimodality, Long Context, and Next Generation Agentic Capabilities}.
\newblock
  \url{https://storage.googleapis.com/deepmind-media/gemini/gemini_v2_5_report.pdf},
  June 2025.

\bibitem{llm-evaluation}
Mingqi Gao, Xinyu Hu, Xunjian Yin, Jie Ruan, Xiao Pu, and Xiaojun Wan.
\newblock Llm-based nlg evaluation: Current status and challenges.
\newblock {\em Computational Linguistics}, pages 1--27, 2025.

\bibitem{google_PromptEng}
{Google Gemini Team}.
\newblock Gemini for google workspace: Prompting guide 101.
\newblock
  \url{https://services.google.com/fh/files/misc/gemini-for-google-workspace-prompting-guide-101.pdf},
  May 2024.

\bibitem{systematic_promptEng}
Pranab Sahoo, Ayush~Kumar Singh, Sriparna Saha, Vinija Jain, Samrat Mondal, and
  Aman Chadha.
\newblock A systematic survey of prompt engineering in large language models:
  Techniques and applications.
\newblock {\em arXiv preprint arXiv:2402.07927}, 2024.

\bibitem{obrien2007caution}
Robert~M. O'Brien.
\newblock A caution regarding rules of thumb for variance inflation factors.
\newblock {\em Quality \& Quantity}, 41(5):673--690, 2007.

\bibitem{krackhardt1988predicting}
David Krackhardt.
\newblock Predicting with networks: Nonparametric multiple regression analysis
  of dyadic data.
\newblock {\em Social networks}, 10(4):359--381, 1988.

\end{thebibliography}
	
\end{document}